\documentclass[]{interact}

\usepackage{epstopdf}
\usepackage[caption=false]{subfig}

\usepackage[numbers,sort&compress]{natbib}
\bibpunct[, ]{[}{]}{,}{n}{,}{,}
\renewcommand\bibfont{\fontsize{10}{12}\selectfont}

\theoremstyle{plain}

\theoremstyle{definition}

\theoremstyle{remark}

\begin{document}

\articletype{ARTICLE TEMPLATE}

\title{Solid-State High-Order Harmonic Generation: Emerging Frontiers in Ultrafast and Quantum Light Science}

\author{
\name{M.~F. Ciappina\textsuperscript{a,b,c}\thanks{CONTACT M.~F. Ciappina. Email: marcelo.ciappina@gtiit.edu.cn}}
\affil{\textsuperscript{a}Department of Physics, Guangdong Technion - Israel Institute of Technology, Shantou, Guangdong, China; \textsuperscript{b}Technion – Israel Institute of Technology, Haifa, Israel; \textsuperscript{c}Guangdong Provincial Key Laboratory of Materials and Technologies for Energy Conversion, Guangdong Technion - Israel Institute of Technology, Shantou, Guangdong, China}
}

\maketitle

\begin{abstract}
High-order harmonic generation (HHG) in solids has emerged as a versatile platform for exploring ultrafast and quantum-coherent phenomena in condensed matter. 
Recent advances reveal Berry-phase and topological effects in harmonic emission, strong-field control of excitons and lattice motion, the generation of nonclassical light states driven by quantum and squeezed fields, and the emergence of orbital-angular-momentum transfer in solid-state high-harmonic generation. Nanostructured and hybrid plasmonic–semiconductor platforms enable enhanced and spectrally tunable HHG, while interferometric and cryogenic setups allow attosecond-resolved phase measurements. On the theoretical side, multiband and topological models incorporating dephasing, propagation, and electron–hole coherence effects have deepened our understanding of the interplay between interband and intraband dynamics. These developments establish solid-state HHG as a bridge between ultrafast spectroscopy, quantum optics, and material science, paving the way toward quantum-engineered attosecond sources and coherent control of light–matter interactions in solids.
\end{abstract}



\begin{figure}[h!]
\centering 
\includegraphics[width=1\linewidth]{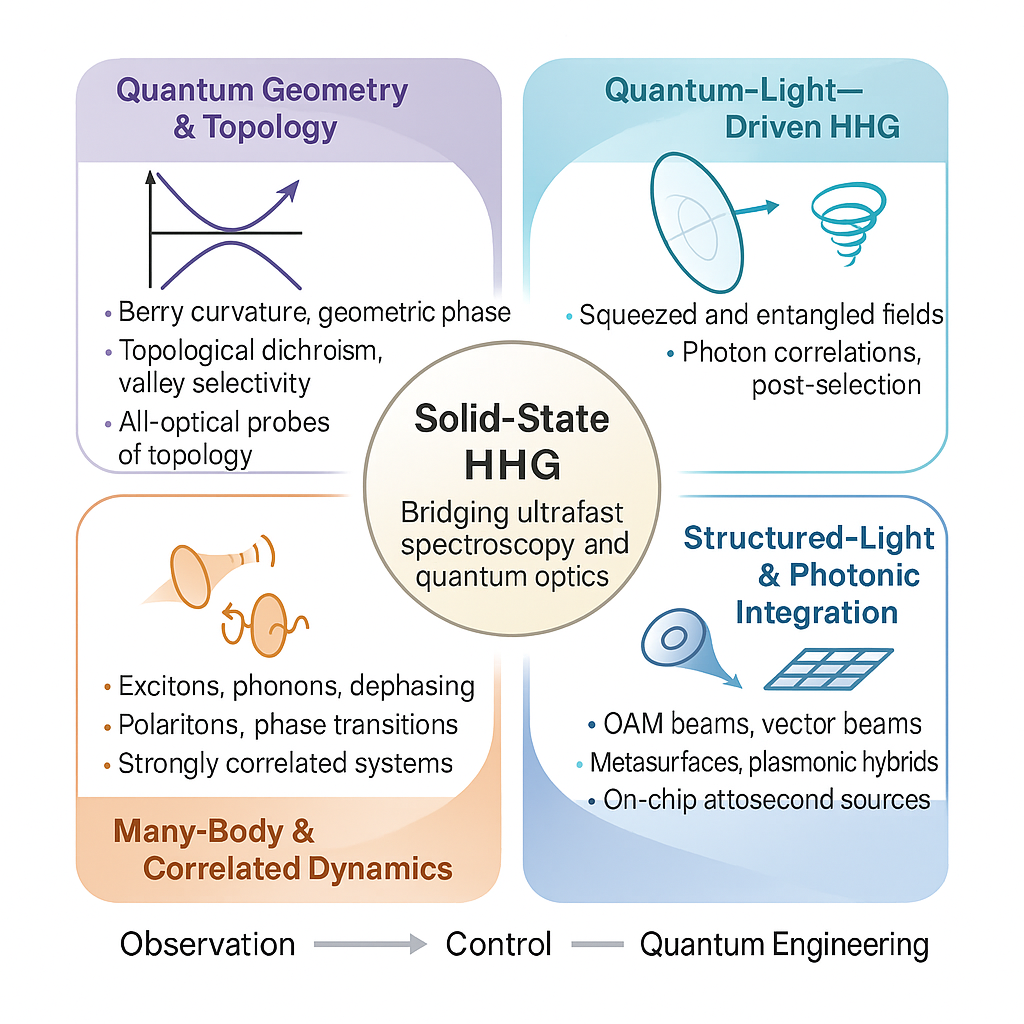}
\caption{Conceptual map of emerging frontiers in solid-state high-harmonic generation (HHG).
Quantum geometry/topology, many-body correlations, and quantum-optical control converge to define a unified framework where HHG acts simultaneously as a probe and generator of quantum-coherent light in solids.}
\label{Fig1}
\end{figure}

Solid-state high-order harmonic generation (HHG) has rapidly evolved into a mature discipline linking ultrafast spectroscopy, quantum optics, and condensed-matter physics. Since the first reports of nonperturbative harmonic emission from bulk crystals, the field has moved beyond proof-of-principle experiments to become a precision platform for exploring electronic structure, topological order, and many-body correlations under extreme fields~\cite{Ciappina_AdvPhysX_7_2022}. We aim to provide a concise update that consolidates previous perspectives while introducing, among other novel testbeds, a quantum-optical dimension, where both the driving and emitted fields exhibit genuinely nonclassical statistics.  The main emerging directions are summarized in Fig.\ref{Fig1}, which illustrates the convergence of topology, correlations, quantum-optical control and structured-light within solid-state HHG.

A major conceptual advance lies in the realization that HHG encodes the quantum geometry of solids. Direct measurement of the interband Berry phase~\cite{Uzan-Narovlansky2024} confirmed that the harmonic polarization and phase depend on Berry curvature, while complementary work~\cite{Garcia-Cabrera2024,Bai_PRL_133_243801_2024}, demonstrated topological dichroism and valley-selective responses in crystalline materials. These studies firmly established HHG as a sensitive probe of geometric and topological band properties, offering an all-optical approach to characterize Berry curvature and orbital magnetic moments. The theoretical foundation for these effects had been clarified earlier~\cite{Ciappina_AdvPhysX_7_2022,Yue_JOSAB_39_535_2022,Kolesik_JOSAB_41_B7_2024}, and the experimental verifications now transform them into a routine diagnostic of quantum geometry.

Parallel progress has occurred in many-body and excitonic regimes. High-harmonic spectroscopy of bound excitons~\cite{Jensen_PRA_109_063104_2024,Heide_NatPhys_20_1546_2024}, and modeling of correlated states~\cite{Yu_PRA_112_2025,Magar_PRB_112_045422_2025,Gnawali_PRB_108_115434_2023}, have shown that interband coherence and dephasing determine both the spectral shape and phase of the harmonics. The inclusion of electron–hole correlation in multiband semiconductor Bloch equations explains the fine interference features and cutoff modulation observed in wide-bandgap and 2D materials~\cite{Heide_Optica_9_512_2022,ZhangM_PRA_112_2025}. The capacity to follow excitonic dynamics in real time positions HHG as a powerful tool to explore light-induced phase transitions and transient polaritonic effects.

Experimentally, phase stability and interferometric precision have reached attosecond-scale resolution. EUV interferometry and Mach–Zehnder phase measurements in solids~\cite{Koll_Optica_12_1606_2025,Uchida_Optica_11_1130_2024} enable reconstruction of the harmonic emission phase, providing access to subcycle dynamics of electrons and lattice motion. Cryogenic HHG platforms~\cite{Kohrell_RSI_95_023906_2024} suppress thermal noise and improve phase coherence, while hybrid plasmonic–semiconductor systems~\cite{Qi_OptLett_50_3050_2025} and metasurface-based enhancement~\cite{Jiang_PRB_110_245424_2024} have achieved significant yield gains. Together, these developments define a new generation of attosecond nanoscopy in solids.

Alternative excitation channels have broadened the HHG landscape. Phonon-assisted photodoping~\cite{Zhang_npJCompMat_10_202_2024}, spin-coupled harmonics in magnetic materials~\cite{Allafi_PRB_110_064420_2024,Kanega_PRB_110_035303_2024,Werner_PRB_110_235101_2024,Yarmohammadi_PRB_110_134442_2024}, and DC-field-driven graphene emission~\cite{Kanega_PRB_110_035303_2024} illustrate how spin, lattice, and charge degrees of freedom couple in extreme nonlinear response. In parallel, confined and edge states in topological nanostructures~\cite{Magar_PRB_112_045422_2025,Gnawali_PRB_108_115434_2023,Liang_PRB_111_085107_2025,Nivash_PRB_112_2025}, give rise to spatially selective emission. The resulting harmonics encode the interplay between topology and confinement, suggesting strategies to control light–matter coupling through nanoscale design.

From a theoretical standpoint, macroscopic propagation and interface modeling have become essential. Kolesik~\cite{Kolesik_JOSAB_41_B7_2024,Kolesik_PRA_110_033512_2024} developed full 3D propagation models connecting microscopic polarization with experimental observables, while systematic studies on substrate materials~\cite{KoricanBarlay_JPCA_128_9337_2024} and thin films~\cite{JourniganBookChap2024} clarified how phase matching and coherence length affect the detected signal. Recent ab initio and many-body approaches have further extended this framework: anomalous high-harmonic generation driven by Berry curvature in monolayer systems~\cite{Yue_PRL_130_166903_2023}, Bloch-wave phase matching linking microscopic interference to macroscopic coherence~\cite{Li_PRL_133_116902_2024}, and resonance-induced amplification of nonlinear response in strongly correlated bosons~\cite{Prakash_PRB_112_085152_2025} illustrate how quantum geometry, lattice structure, and correlations collectively determine harmonic yield and phase. Together, these models reconcile discrepancies between experiments on bulk and thin-film systems, bringing solid HHG modeling to a level of predictive power comparable to gas-phase theories.

An equally transformative direction is the rise of quantum-optical strong-field physics. Experiments with squeezed-vacuum driving fields~\cite{Rasputnyi_NatPhys_2024} have revealed nonclassical photon statistics in harmonic emission. Quantum-field theoretical models~\cite{Gothelf_PRA_111_063105_2025,LiJ_PRA_112_033507_2025,RiveraDean_PRB_109_035203_2024,RiveraPost2025,LangePRL2025,Lemieux2025,Ciappina2025}, show that quantum correlations between photons and electrons modify the harmonic spectrum, leading to antibunching, squeezing, and entanglement features in the emitted light. These findings blur the traditional boundary between HHG and quantum optics, positioning solid HHG as a tabletop source of engineered quantum light.

The emergence of quantum-light–driven HHG signals a conceptual convergence between attosecond science and quantum information. Theoretical extensions~\cite{Gothelf_PRA_111_063105_2025,LiJ_PRA_112_033507_2025,RiveraDean_PRB_109_035203_2024,RiveraPost2025,LangePRL2025} predict that squeezed or entangled driving fields can imprint quantum correlations on harmonic photons, generating hybrid exciton–photon states and cat-like superpositions. These predictions align with experimental observations of bright squeezed-vacuum-driven emission~\cite{Rasputnyi_NatPhys_2024} and nonclassical photon statistics in semiconductor HHG~\cite{RiveraDean_PRB_109_035203_2024}. The control of harmonic emission through quantum-state engineering represents one of the most promising frontiers of strong-field physics, enabling tunable quantum-optical nonlinearities in bulk media.

Progress in laser technology has paralleled these conceptual advances. Fiber-based and mid-IR laser systems with excellent phase stability~\cite{BoukhaouiAPLPhotonics2025} now drive HHG in solids at high repetition rates and modest intensities. Two-color, non-collinear mixing~\cite{RoscamAbbing_NatCommun_15_10119_2024} and all-optical modulation~\cite{vanEssen_ACSPhot2024} enable precise control over emission amplitude and polarization. Phase-resolved interferometry~\cite{Koll_Optica_12_1606_2025,Uchida_Optica_11_1130_2024} allows the retrieval of the emission phase with attosecond precision, establishing HHG as a quantitative probe of ultrafast electron and lattice motion. Together, these tools are transforming HHG into a metrological technique for condensed-matter dynamics.

New classes of materials and nanostructures have further diversified the field. Moiré superlattices exhibit strong many-body enhancement of HHG~\cite{ChangLee2024}, while flat-band two-dimensional materials~\cite{LiF_PRB_111_045402_2025} show interband-polarization-dominated emission linked to carrier localization. Engineered dimerized chains~\cite{Nivash_PRB_112_2025} and Kitaev lattices reveal how structural modulation shapes harmonic polarization and cutoff scaling. Studies on large-bandgap substrates~\cite{KoricanBarlay_JPCA_128_9337_2024}  and thin-film architectures~\cite{JourniganBookChap2024} confirm that interface engineering and coherence-length optimization are decisive for macroscopic yield.

Integration with nanophotonics provides unprecedented control over near-field enhancement and spectral tunability. Hybrid plasmonic–semiconductor systems~\cite{Qi_OptLett_50_3050_2025} and nonlocal metasurfaces~\cite{Jiang_PRB_110_245424_2024} employ quasiguided modes for quasi-phase matching, producing order-of-magnitude yield increases. Dielectric metasurfaces allow precise phase engineering without thermal degradation, while cryogenic and closed-cycle configurations~\cite{Kohrell_RSI_95_023906_2024} maintain stability. These advances establish solid HHG as a compact, scalable, and chip-compatible coherent light source.

Temporal control has also progressed rapidly. Subcycle field manipulation~\cite{Gindl_PRL_134_176903_2025}  enables attosecond shaping of the emitted radiation, while theoretical analyses~\cite{GoulielmakisBrabec_NatPhoton_16_411_2022,Hirori_JPCL_15_2024} unify the description of harmonic emission and ultrafast current control in solids. Attosecond pulse formation, phase locking, and carrier-envelope control now link HHG directly to petahertz electronics, where light fields drive electronic motion on subfemtosecond scales. Experiments resolving lattice vibrations in harmonic emission~\cite{Zhang_NatPhoton_18_792_2024} reveal the coupling between phonons and nonlinear polarization, completing the picture of a fully correlated electron-lattice response.

The field has also expanded conceptually toward topological and correlated materials. High harmonics from topological insulators and altermagnets~\cite{Allafi_PRB_110_064420_2024,Kanega_PRB_110_035303_2024,Werner_PRB_110_235101_2024,Liang_PRB_111_085107_2025} reveal how symmetry breaking and spin texture affect emission polarization. HHG from confined systems such as graphene quantum dots~\cite{Magar_PRB_112_045422_2025,Gnawali_PRB_108_115434_2023} and magnetic heterostructures demonstrates sensitivity to edge topology and spin order. Such findings suggest that HHG can serve as an all-optical probe of topological invariants and correlated magnetism under nonequilibrium conditions.

In parallel, extensive simulation efforts~\cite{Yue_JOSAB_39_535_2022,Kolesik_JOSAB_41_B7_2024} have standardized computational tools and clarified the roles of interband and intraband mechanisms, anisotropy, and dephasing. Macroscopic simulations now routinely combine semiconductor Bloch equations with full-field propagation~\cite{Kolesik_PRA_110_033512_2024}, achieving quantitative agreement with experiments. These models predict subtle features such as spectral self-compression and orientation-dependent cutoff modulation, which have been confirmed experimentally~\cite{ZhangM_PRA_112_2025,KoricanBarlay_JPCA_128_9337_2024}. The result is a unified understanding linking microscopic electron trajectories with macroscopic harmonic emission.

The generation of high harmonics carrying orbital angular momentum (OAM) has evolved from initial demonstrations of OAM conservation in solids to the full control of structured attosecond emission. Early studies established that when a Laguerre–Gaussian beam drives HHG in a semiconductor crystal, the harmonics inherit a topological charge scaling linearly with the harmonic order, confirming angular-momentum conservation and enabling on-chip diffractive control of vortex harmonics~\cite{OptLett2019}. Subsequent advances extended this concept to the extreme-ultraviolet domain, achieving very high topological charges and complete wavefront characterization through high-resolution sensing, which revealed the sensitivity of the harmonic phase structure to the purity of the driving mode~\cite{ACSPhotonics2022}. More recently, a symmetry-based framework unified the spin and orbital angular momentum of light in anisotropic solids, showing that the total angular momentum is conserved through multiscale dynamical symmetries that couple crystal anisotropy with spin–orbit interactions~\cite{SciAdv2024}. Topological analyses further demonstrated that in crystalline and two-dimensional materials, the harmonic field can encode microscopic information about the nonlinear response and symmetry breaking of the medium~\cite{Garcia-Cabrera2024}. Theoretical developments combining semiconductor Bloch equations with thin-slab models have reproduced these observations and predicted that the coherent synthesis of multiple harmonics can produce attosecond vortex pulses, linking structured-light HHG with ultrafast quantum optics and establishing solids as versatile platforms for quantum-engineered twisted attosecond sources~\cite{UScience2025,JPhysD2025}.

Recent advances have marked a transition from merely observing high-harmonic generation (HHG) to actively controlling it through material design, quantum statistics, and photonic integration. This evolution highlights the growing role of post-selection and squeezed-light techniques in producing nonclassical harmonics, expanding HHG’s potential for ultrafast quantum sensing and coherent XUV photonics. At the same time, the emergence of structured-light HHG has opened new possibilities for controlling the angular momentum and topology of light–matter interactions, as harmonics driven by vortex or vector beams have been shown to conserve and reshape orbital angular momentum, enabling tailored spin–orbit couplings and topological field structures within solids. The discovery of multiscale dynamical symmetries in anisotropic media has further unified the spin and orbital degrees of freedom under a single conservation rule, while theoretical models based on semiconductor Bloch dynamics now predict the generation of attosecond vortex pulse trains through the coherent synthesis of multiple harmonics. Collectively, these developments signify that solid-state HHG has reached a stage of conceptual maturity: it now functions both as a probe and a generator, revealing the internal dynamics of electrons, excitons, and phonons while emitting coherent and potentially quantum-correlated light. The interplay of topology, correlations, and quantum optics thus defines a fertile arena for exploration, as HHG in solids transforms from a nonlinear optical curiosity into a foundational instrument of attosecond science and quantum technology—bridging the microscopic physics of materials with the macroscopic control of light.

\section*{Disclosure statement}

No potential conflict of interest was reported by the author(s).


\section*{Funding}

M.~F.~C.~acknowledges support by the National Key Research and Development Program of China (Grant No.~2023YFA1407100), Guangdong Province Science and Technology Major Project (Future functional materials under extreme conditions - 2021B0301030005), the Guangdong Natural Science Foundation (General Program project No. 2023A1515010871), and the National Natural Science Foundation of China (Grant No. 12574092).








\bibliographystyle{tfq}
\bibliography{review}

\begin{thebibliography}{10}
\newcommand{\printfirst}[2]{#1}
\newcommand{\switchargs}[2]{#2#1}
\providecommand{\url}[1]{\normalfont{#1}}
\providecommand{\urlprefix}{Available at }

\bibitem{Ciappina_AdvPhysX_7_2022}
J. Park, A. Subramani, S. Kim, and M.F. Ciappina, \emph{Recent trends in
  high-order harmonic generation in solids}, Adv. Phys.: X 7 (2022), p.
  2003244, \urlprefix\url{https://doi.org/10.1080/23746149.2021.2003244}.

\bibitem{Uzan-Narovlansky2024}
A.J. Uzan-Narovlansky, L. Faeyrman, G.G. Brown, S. Shames, V. Narovlansky, J.
  Xiao, T. Arusi-Parpar, O. Kneller, B.D. Bruner, O. Smirnova, R.E.F. Silva, B.
  Yan, {\'A}. Jim{\'e}nez-Gal{\'a}n, M. Ivanov, and N. Dudovich,
  \emph{Observation of interband berry phase in laser-driven crystals}, Nature
  626 (2024), pp. 66--71,
  \urlprefix\url{https://doi.org/10.1038/s41586-023-06828-5}.

\bibitem{Garcia-Cabrera2024}
A. Garc{\'i}a-Cabrera, R. Boyero-Garc{\'i}a, {\'O}. Zurr{\'o}n-Cifuentes, J.
  Serrano, J.S. Rom{\'a}n, L. Plaja, and C. Hern{\'a}ndez-Garc{\'i}a,
  \emph{Topological high-harmonic spectroscopy}, Commun. Phys. 7 (2024), p.~28,
  \urlprefix\url{https://doi.org/10.1038/s42005-023-01511-7}.

\bibitem{Bai_PRL_133_243801_2024}
Y. Bai, Y. Jiang, W. Zheng, H. Xu, J. Chen, S. Wang, C. Liu, R. Li, and P. Liu,
  \emph{{Probing Berry Phase Effect in Topological Surface States}}, Phys. Rev.
  Lett. 133 (2024), p. 243801,
  \urlprefix\url{https://link.aps.org/doi/10.1103/PhysRevLett.133.243801}.

\bibitem{Yue_JOSAB_39_535_2022}
L. Yue and M.B. Gaarde, \emph{Introduction to theory of high-harmonic
  generation in solids: tutorial}, J. Opt. Soc. Am. B 39 (2022), pp. 535--555,
  \urlprefix\url{https://opg.optica.org/josab/abstract.cfm?URI=josab-39-2-535}.

\bibitem{Kolesik_JOSAB_41_B7_2024}
M. Kolesik, \emph{Sample-orientation effects in solid-state high-harmonic
  generation: computational study of {GaAs}}, J. Opt. Soc. Am. B 41 (2024), pp.
  B7--B13,
  \urlprefix\url{https://opg.optica.org/josab/abstract.cfm?URI=josab-41-6-B7}.

\bibitem{Jensen_PRA_109_063104_2024}
S.V.B. Jensen, L.B. Madsen, A. Rubio, and N. Tancogne-Dejean,
  \emph{High-harmonic spectroscopy of strongly bound excitons in solids}, Phys.
  Rev. A 109 (2024), p. 063104,
  \urlprefix\url{https://link.aps.org/doi/10.1103/PhysRevA.109.063104}.

\bibitem{Heide_NatPhys_20_1546_2024}
C. Heide, Y. Kobayashi, S.R.U. Haque, and S. Ghimire, \emph{Ultrafast
  high-harmonic spectroscopy of solids}, Nat. Phys. 20 (2024), pp. 1546--1557,
  \urlprefix\url{https://doi.org/10.1038/s41567-024-02640-8}.

\bibitem{Yu_PRA_112_2025}
C. Yu, \emph{Incorporating dephasing effects into simulations of harmonic
  generation from finite-size topological insulators}, Phys. Rev. A 112 (2025),
  p. 033112, \urlprefix\url{https://link.aps.org/doi/10.1103/zggs-4v8m}.

\bibitem{Magar_PRB_112_045422_2025}
K.R. Magar and V. Apalkov, \emph{High harmonic generation in {Haldane} model
  quantum dots}, Phys. Rev. B 112 (2025), p. 045422,
  \urlprefix\url{https://link.aps.org/doi/10.1103/65ql-ytqw}.

\bibitem{Gnawali_PRB_108_115434_2023}
S. Gnawali and V. Apalkov, \emph{High harmonic generation governed by edge
  states in triangular graphene quantum dots}, Phys. Rev. B 108 (2023), p.
  115434, \urlprefix\url{https://link.aps.org/doi/10.1103/PhysRevB.108.115434}.

\bibitem{Heide_Optica_9_512_2022}
C. Heide, Y. Kobayashi, A.C. Johnson, F. Liu, T.F. Heinz, D.A. Reis, and S.
  Ghimire, \emph{Probing electron-hole coherence in strongly driven {2D}
  materials using high-harmonic generation}, Optica 9 (2022), pp. 512--516,
  \urlprefix\url{https://opg.optica.org/optica/abstract.cfm?URI=optica-9-5-512}.

\bibitem{ZhangM_PRA_112_2025}
M. Zhang, Y. Zhang, J. Li, L. Hu, R. Zhong, S. Yue, and H. Du, \emph{Impact of
  excitation mechanism on ellipticity and orientation dependence of solid
  harmonic yield}, Phys. Rev. A 112 (2025), p. 033534,
  \urlprefix\url{https://link.aps.org/doi/10.1103/lvp4-w3kb}.

\bibitem{Koll_Optica_12_1606_2025}
L.M. Koll, S.V.B. Jensen, P.J. van  Essen, B. de  Keijzer, E. Olsson, J.
  Cottom, T. Witting, A. Husakou, M.J.J. Vrakking, L.B. Madsen, P.M. Kraus, and
  P. J\"{u}rgens, \emph{Extreme ultraviolet high-harmonic interferometry of
  excitation-induced bandgap dynamics in solids}, Optica 12 (2025), pp.
  1606--1614,
  \urlprefix\url{https://opg.optica.org/optica/abstract.cfm?URI=optica-12-10-1606}.

\bibitem{Uchida_Optica_11_1130_2024}
K. Uchida and K. Tanaka, \emph{High harmonic {Mach--Zehnder} interferometer for
  probing sub-laser-cycle electron dynamics in solids}, Optica 11 (2024), pp.
  1130--1137,
  \urlprefix\url{https://opg.optica.org/optica/abstract.cfm?URI=optica-11-8-1130}.

\bibitem{Kohrell_RSI_95_023906_2024}
F. Kohrell, B.R. Nebgen, J.A. Spies, R. Hollinger, A. Zong, C. Uzundal, C.
  Spielmann, and M. Zuerch, \emph{A solid-state high harmonic generation
  spectrometer with cryogenic cooling}, Rev. Sci. Instrum. 95 (2024), p.
  023906, \urlprefix\url{https://doi.org/10.1063/5.0174407}.

\bibitem{Qi_OptLett_50_3050_2025}
Z. Wan, Q. Li, H. Yang, L. He, L. Lu, Z.Y. Chen, Q. Zhang, Y. Hou, L. Wu, and
  H. Liang, \emph{High harmonic generation in solids with an ultra-low
  threshold of 7 {GW/cm}$^{2}$}, Opt. Lett. 50 (2025), pp. 3050--3053,
  \urlprefix\url{https://opg.optica.org/ol/abstract.cfm?URI=ol-50-9-3050}.

\bibitem{Jiang_PRB_110_245424_2024}
H. Jiang, Y. Cai, and Z. Han, \emph{Enhanced high-harmonic generation with
  spectral tunability in nonlocal metasurfaces enabled by the excitation of
  quasiguided modes}, Phys. Rev. B 110 (2024), p. 245424,
  \urlprefix\url{https://link.aps.org/doi/10.1103/PhysRevB.110.245424}.

\bibitem{Zhang_npJCompMat_10_202_2024}
J. Zhang, O. Neufeld, N. Tancogne-Dejean, I.T. Lu, H. H{\"u}bener, U.
  De~Giovannini, and A. Rubio, \emph{Enhanced high harmonic efficiency through
  phonon-assisted photodoping effect}, npj Comput. Mater. 10 (2024), p. 202,
  \urlprefix\url{https://doi.org/10.1038/s41524-024-01399-z}.

\bibitem{Allafi_PRB_110_064420_2024}
N.M. Allafi, M.H. Kolodrubetz, M. Bukov, V. Oganesyan, and M. Yarmohammadi,
  \emph{Spin high harmonic generation through terahertz laser-driven phonons},
  Phys. Rev. B 110 (2024), p. 064420,
  \urlprefix\url{https://link.aps.org/doi/10.1103/PhysRevB.110.064420}.

\bibitem{Kanega_PRB_110_035303_2024}
M. Kanega and M. Sato, \emph{High-harmonic generation in graphene under the
  application of a dc electric current: From perturbative to nonperturbative
  regime}, Phys. Rev. B 110 (2024), p. 035303,
  \urlprefix\url{https://link.aps.org/doi/10.1103/PhysRevB.110.035303}.

\bibitem{Werner_PRB_110_235101_2024}
P. Werner, M. Lysne, and Y. Murakami, \emph{High harmonic generation in
  altermagnets}, Phys. Rev. B 110 (2024), p. 235101,
  \urlprefix\url{https://link.aps.org/doi/10.1103/PhysRevB.110.235101}.

\bibitem{Yarmohammadi_PRB_110_134442_2024}
M. Yarmohammadi and M.H. Kolodrubetz, \emph{Terahertz high-harmonic generation
  in gapped antiferromagnetic chains}, Phys. Rev. B 110 (2024), p. 134442,
  \urlprefix\url{https://link.aps.org/doi/10.1103/PhysRevB.110.134442}.

\bibitem{Liang_PRB_111_085107_2025}
S. Liang, L. Li, J. Li, S. Qiao, Y. Zhang, W. He, D. Wu, X. Zhu, P. Lan, and P.
  Lu, \emph{Circular dichroism of the dipole transition in nanosized
  topological insulators}, Phys. Rev. B 111 (2025), p. 085107,
  \urlprefix\url{https://link.aps.org/doi/10.1103/PhysRevB.111.085107}.

\bibitem{Nivash_PRB_112_2025}
R. Nivash, S. Srinidhi, J.N. Bandyopadhyay, and A.R. Holkundkar,
  \emph{Engineering harmonic emission through spatial modulation in a kitaev
  chain}, Phys. Rev. B 112 (2025), p. 094311,
  \urlprefix\url{https://link.aps.org/doi/10.1103/nf6m-1vfh}.

\bibitem{Kolesik_PRA_110_033512_2024}
M. Kolesik, \emph{Propagation and material-interface effects in the
  higher-order harmonic radiation from solid-state samples}, Phys. Rev. A 110
  (2024), p. 033512,
  \urlprefix\url{https://link.aps.org/doi/10.1103/PhysRevA.110.033512}.

\bibitem{KoricanBarlay_JPCA_128_9337_2024}
E. Korican-Barlay, B.R. Nebgen, J.A. Spies, and M.W. Zuerch, \emph{Solid-state
  high harmonic generation in common large bandgap substrate materials}, J.
  Phys. Chem. A. 128 (2024), pp. 9337--9344,
  \urlprefix\url{https://doi.org/10.1021/acs.jpca.4c04991}.

\bibitem{JourniganBookChap2024}
T.D. Journigan and M. Chini, \emph{Chapter one - high-order harmonic generation
  in thin-film solids}, in \emph{Advances in Atomic, Molecular, and Optical
  Physics}, L.F. DiMauro, H. Perrin, and S. Yelin, eds., Advances In Atomic,
  Molecular, and Optical Physics Vol.~73, Academic Press,  2024, pp. 1--31,
  \urlprefix\url{https://www.sciencedirect.com/science/article/pii/S1049250X24000016}.

\bibitem{Yue_PRL_130_166903_2023}
L. Yue and M.B. Gaarde, \emph{{Characterizing Anomalous High-Harmonic
  Generation in Solids}}, Phys. Rev. Lett. 130 (2023), p. 166903,
  \urlprefix\url{https://link.aps.org/doi/10.1103/PhysRevLett.130.166903}.

\bibitem{Li_PRL_133_116902_2024}
L. Li, Y. Zhang, J. Li, P. Lan, and P. Lu, \emph{{Bloch-Wave Phase Matching of
  High Harmonic Generation in Solids}}, Phys. Rev. Lett. 133 (2024), p. 116902,
  \urlprefix\url{https://link.aps.org/doi/10.1103/PhysRevLett.133.116902}.

\bibitem{Prakash_PRB_112_085152_2025}
A. Prakash, D. Dutta, A. Roy, and K. Saha, \emph{Frequency-selective
  amplification of nonlinear response in strongly correlated bosons}, Phys.
  Rev. B 112 (2025), p. 085152,
  \urlprefix\url{https://link.aps.org/doi/10.1103/2x73-gp8k}.

\bibitem{Rasputnyi_NatPhys_2024}
A. Rasputnyi, Z. Chen, M. Birk, O. Cohen, I. Kaminer, M. Kr{\"u}ger, D.
  Seletskiy, M. Chekhova, and F. Tani, \emph{High-harmonic generation by a
  bright squeezed vacuum}, Nat. Phys. 20 (2024), pp. 1960--1965,
  \urlprefix\url{https://doi.org/10.1038/s41567-024-02659-x}.

\bibitem{Gothelf_PRA_111_063105_2025}
R.V. Gothelf, C.S. Lange, and L.B. Madsen, \emph{High-order harmonic generation
  in a crystal driven by quantum light}, Phys. Rev. A 111 (2025), p. 063105,
  \urlprefix\url{https://link.aps.org/doi/10.1103/PhysRevA.111.063105}.

\bibitem{LiJ_PRA_112_033507_2025}
J. Li, Z. Lyu, H. Liu, and Y. Liu, \emph{Quantum-optical signatures of
  solid-state high-harmonic generation driven by squeezed coherent states},
  Phys. Rev. A 112 (2025), p. 033507,
  \urlprefix\url{https://link.aps.org/doi/10.1103/vmk1-qjpj}.

\bibitem{RiveraDean_PRB_109_035203_2024}
J. Rivera-Dean, P. Stammer, A.S. Maxwell, T. Lamprou, A.F. Ord\'o\~nez, E.
  Pisanty, P. Tzallas, M. Lewenstein, and M.F. Ciappina, \emph{Nonclassical
  states of light after high-harmonic generation in semiconductors: A
  bloch-based perspective}, Phys. Rev. B 109 (2024), p. 035203,
  \urlprefix\url{https://link.aps.org/doi/10.1103/PhysRevB.109.035203}.

\bibitem{RiveraPost2025}
J. Rivera-Dean, T. Lamprou, E. Pisanty, M.F. Ciappina, P. Tzallas, M.
  Lewenstein, and P. Stammer, \emph{Quantum state engineering of light using
  intensity measurements and postselection}, Phys. Rev. A 112 (2025), p.
  013110, \urlprefix\url{https://link.aps.org/doi/10.1103/11vz-9gcz}.

\bibitem{LangePRL2025}
C.S. Lange, T. Hansen, and L.B. Madsen, \emph{Excitonic enhancement of squeezed
  light in quantum-optical high-harmonic generation from a mott insulator},
  Phys. Rev. Lett. 135 (2025), p. 043603,
  \urlprefix\url{https://link.aps.org/doi/10.1103/wyk5-k8tk}.

\bibitem{Lemieux2025}
S. Lemieux, S.A. Jalil, D.N. Purschke, N. Boroumand, T.J. Hammond, D.
  Villeneuve, A. Naumov, T. Brabec, and G. Vampa, \emph{Photon bunching in
  high-harmonic emission controlled by quantum light}, Nat. Photonics 19
  (2025), pp. 767--771,
  \urlprefix\url{https://doi.org/10.1038/s41566-025-01673-6}.

\bibitem{Ciappina2025}
M. Ciappina, \emph{High-harmonic generation meets quantum optics}, Nat.
  Photonics 19 (2025), pp. 666--667,
  \urlprefix\url{https://doi.org/10.1038/s41566-025-01702-4}.

\bibitem{BoukhaouiAPLPhotonics2025}
D. Boukhaoui, A. Mikhneva, S. Idlahcen, J. Houard, T. Godin, L. Guiramand, I.
  Blum, F. Amrani, F. Gérôme, F. Benabid, D. Gauthier, W. Boutu, H. Merdji,
  A. Vella, and A. Hideur, \emph{High-harmonic generation in solids from a
  high-energy fiber laser system}, APL Photonics 10 (2025), p. 026106,
  \urlprefix\url{https://doi.org/10.1063/5.0244415}.

\bibitem{RoscamAbbing_NatCommun_15_10119_2024}
S.D.C. Roscam~Abbing, N. Kuzkova, R. van~der  Linden, F. Campi, B. de  Keijzer,
  C. Morice, Z.Y. Zhang, M.L.S. van~der  Geest, and P.M. Kraus, \emph{Enhancing
  the efficiency of high-order harmonics with two-color non-collinear wave
  mixing in silica}, Nat. Commun. 15 (2024), p. 8335,
  \urlprefix\url{https://doi.org/10.1038/s41467-024-52774-9}.

\bibitem{vanEssen_ACSPhot2024}
P.J. van  Essen, Z. Nie, B. de  Keijzer, and P.M. Kraus, \emph{Toward complete
  all-optical intensity modulation of high-harmonic generation from solids},
  ACS Photonics 11 (2024), pp. 1832--1843,
  \urlprefix\url{https://doi.org/10.1021/acsphotonics.4c00156}.

\bibitem{ChangLee2024}
V. Chang~Lee, L. Yue, M.B. Gaarde, Y.h. Chan, and D.Y. Qiu, \emph{Many-body
  enhancement of high-harmonic generation in monolayer {MoS$_2$}}, Nat. Commun.
  15 (2024), p. 6228,
  \urlprefix\url{https://doi.org/10.1038/s41467-024-50534-3}.

\bibitem{LiF_PRB_111_045402_2025}
F. Li, C. Qian, X. Bo, S. Li, T. Wu, C. Yu, Y. Wang, H. Wu, and R. Lu,
  \emph{Interband-polarization dominated high harmonic generation from
  flat-band $\mathrm{NbOC}{\mathrm{l}}_{2}$ materials in strong laser fields},
  Phys. Rev. B 111 (2025), p. 045402,
  \urlprefix\url{https://link.aps.org/doi/10.1103/PhysRevB.111.045402}.

\bibitem{Gindl_PRL_134_176903_2025}
A. Gindl, P. Suthar, F.c.v. Troj\'anek, P. Mal\'y, T.J.Y. Derrien, and M.
  Koz\'ak, \emph{Attosecond control of solid-state high harmonic generation
  using $\ensuremath{\omega}\text{\ensuremath{-}}3\ensuremath{\omega}$ fields},
  Phys. Rev. Lett. 134 (2025), p. 176903,
  \urlprefix\url{https://link.aps.org/doi/10.1103/PhysRevLett.134.176903}.

\bibitem{GoulielmakisBrabec_NatPhoton_16_411_2022}
E. Goulielmakis and T. Brabec, \emph{High harmonic generation in condensed
  matter}, Nat. Photonics 16 (2022), pp. 411--421,
  \urlprefix\url{https://doi.org/10.1038/s41566-022-00988-y}.

\bibitem{Hirori_JPCL_15_2024}
H. Hirori, S.A. Sato, and Y. Kanemitsu, \emph{High-order harmonic generation in
  solids: The role of intraband transitions in extreme nonlinear optics}, J.
  Phys. Chem. Lett. 15 (2024), pp. 2184--2192,
  \urlprefix\url{https://doi.org/10.1021/acs.jpclett.3c03415}, pMID: 38373145.

\bibitem{Zhang_NatPhoton_18_792_2024}
J. Zhang, Z. Wang, F. Lengers, D. Wigger, D.E. Reiter, T. Kuhn, H.J.
  W{\"o}rner, and T.T. Luu, \emph{High-harmonic spectroscopy probes lattice
  dynamics}, Nat. Photonics 18 (2024), pp. 792--798,
  \urlprefix\url{https://doi.org/10.1038/s41566-024-01457-4}.

\bibitem{OptLett2019}
D. Gauthier, S. Kaassamani, D. Franz, R. Nicolas, J.T. Gomes, L. Lavoute, D.
  Gaponov, S. F\'{e}vrier, G. Jargot, M. Hanna, W. Boutu, and H. Merdji,
  \emph{Orbital angular momentum from semiconductor high-order harmonics}, Opt.
  Lett. 44 (2019), pp. 546--549,
  \urlprefix\url{https://opg.optica.org/ol/abstract.cfm?URI=ol-44-3-546}.

\bibitem{ACSPhotonics2022}
A.K. Pandey, A. de~las  Heras, T. Larrieu, J. San~Román, J. Serrano, L. Plaja,
  E. Baynard, M. Pittman, G. Dovillaire, S. Kazamias, C. Hernández-García,
  and O. Guilbaud, \emph{Characterization of extreme ultraviolet vortex beams
  with a very high topological charge}, ACS Photonics 9 (2022), pp. 944--951,
  \urlprefix\url{https://doi.org/10.1021/acsphotonics.1c01768}.

\bibitem{SciAdv2024}
K. Nagai, T. Okamoto, Y. Shinohara, H. Sanada, and K. Oguri,
  \emph{High-harmonic spin-orbit angular momentum generation in crystalline
  solids preserving multiscale dynamical symmetry}, Sci. Adv. 10 (2024), p.
  eado7315,
  \urlprefix\url{https://www.science.org/doi/abs/10.1126/sciadv.ado7315}.

\bibitem{UScience2025}
C. Granados, B.K. Das, C. Heide, S. Ghimire, and M.F. Ciappina, \emph{Toward
  attosecond vortices in semiconductor materials}, Ultrafast Sci. 5 (2025), p.
  0100,
  \urlprefix\url{https://spj.science.org/doi/abs/10.34133/ultrafastscience.0100}.

\bibitem{JPhysD2025}
C. Granados, B.K. Das, W. Gao, and M.F. Ciappina, \emph{{High-order harmonic
  generation in ZnO driven by a Laguerre–Gauss vortex beam}}, J. Phys. D 58
  (2025), p. 365104, \urlprefix\url{https://doi.org/10.1088/1361-6463/adfdf4}.

\end{thebibliography}

\end{document}